\documentclass[
 aip,
 amsmath,amssymb,
 reprint,
]{revtex4-2}

\usepackage{graphicx}
\usepackage{dcolumn}
\usepackage{bm}

\usepackage[utf8]{inputenc}
\usepackage[T1]{fontenc}
\usepackage{mathptmx, makecell}
\usepackage{etoolbox}

\usepackage[colorlinks=True]{hyperref}

\newcommand{\deriv}{\mathrm{d}}

\newcommand*{\rttensor}[1]{\bar{\bar{#1}}}

\newcommand{\Fermi}{\textsc{f}}

\newcommand{\veck}{\mathbf{k}}
\newcommand{\vecv}{\mathbf{v}}

\makeatletter
\def\@email#1#2{%
 \endgroup
 \patchcmd{\titleblock@produce}
  {\frontmatter@RRAPformat}
  {\frontmatter@RRAPformat{\produce@RRAP{*#1\href{mailto:#2}{#2}}}\frontmatter@RRAPformat}
  {}{}
}%
\makeatother
\begin{document}

	
\title{Exploring binary intermetallics for advanced interconnect applications using \textit{ab initio} simulations}

\author{Kiroubanand Sankaran}
\email[]{sankaran@imec.be}
\author{Kristof Moors}
\author{Jean-Philippe Souli\'e}
\author{Christoph Adelmann}
\author{Geoffrey Pourtois}
\affiliation{IMEC, Kapeldreef 75, B-3001 Leuven, Belgium}

\date{\today}

\begin{abstract}
The challenge of increasing copper (Cu) resistivity with diminishing Cu interconnect dimensions in complementary metal-oxide-semiconductor (CMOS) transistors, along with the imperative for efficient electron transport paths to fulfill scaling requirements in interconnects is significant. First-principles electronic structures calculations based on density functional theory have been performed to evaluate the potential scalability of some Cu, Al, Ru and Mo based binary alloys to replace Cu. 
We evaluate the expected sensitivity of the resistivity of these binary alloys to reduced line dimensions with a figure of merit that is based on generalized finite-temperature transport tensors. These transport tensors allow for a straightforward comparison between highly anisotropic intermetallics with given transport directions and Cu, and are evaluated together with their resistance to electromigration. Based on the figure-of-merit analysis, we identify several aluminides that show potential to outperform Cu at reduced interconnect dimensions in terms of their electronic transport and reliability properties.
\end{abstract}
\maketitle

\section{\label{sec:introduction}Introduction}
For over two decades Cu has been used as the main interconnect conductor material to enable the downscaling of electronic circuits in complementary metal-oxide-semiconductor (CMOS) technology.
This continuous aggressive reduction of the physical dimensions of Cu lines is strongly impacting the interconnect electrical resistance and hence the reliability of the overall CMOS circuitry due to electron scattering events induced within the Cu vias by grain boundaries and surface roughness.~\cite{Kapur2002, Clarke2014, Croes2018, Baklanov2014, Lin2018, Nogami2019}
In this context, identifying alternative conductors to Cu is critical to overcome the above-mentioned hurdles.
Contenders must be resilient to electromigration (EM) and diffusion into dielectrics in combination to display smaller increase in resistivity than Cu in tight-pitched narrow lines. In addition, any alternative conductor should also meet other crucial requirements such as excellent thermal conductivity, etchability, barrier-less deposition and to enable cost efficient integration schemes.~\cite{Kapur2002, Clarke2014, Croes2018, Baklanov2014, Lin2018, Nogami2019} This set of constraints makes that the search for alternative metals to Cu has become very challenging.
To advance beyond the copper era, efficient methodologies for funneling materials are essential to identify candidates that meet all necessary requirements.
In this context, combining semiclassical electron transport with first-principles simulations provides an effective approach for this identification.
The semiclassical electron transport models developed by Fuchs \& Sondheimer~\cite{Fuchs1938, Sondheimer1952}and Mayadas \& Shatzkes~\cite{Mayadas1970} suggest that the resistivity of polycrystalline films depends essentially on the film thickness and grain size with a constant of proportionality given by the product of bulk resistivity $\rho_0$ and bulk electron mean free path (MFP) $\lambda$ ($\rho_0\times\lambda$).~\cite{Gall2020}
Metals with short MFP are inherently less affected by less affected by scattering events with characteristic length scale determined by the confinement dimensions,, e.g. surface or grain-boundary scattering with characteristic length scale determined by the confinement dimensions (d) being significantly lower than $\lambda$. That makes them promising for good scalability in resistivity.
However, the downside is that these metals also experience numerous scattering events within the material bulk, leading to high intrinsic bulk resistivity.
Therefore, a balance must be struck between scalability potential ($\rho_0 \times \lambda$) and intrinsic bulk resistivity ($\rho_0$).
Additionally, conductors with higher cohesive/bonding energy than copper (Cu) are expected to be more resilient to electromigration (EM) and diffusion issues into dielectrics, potentially enabling barrier-less interconnect concepts.~\cite{Sankaran2014}

To address these issues, significant research efforts have been dedicated to identifying metal conductor candidates with both short MFP and high cohesive energy.
This approach has recently been recognized as an effective ranking method for evaluating new alternative conductors. Potential candidates to replace Cu range from elemental metals~\cite{Gall2016, Dutta2017} to compounds, including ternary.~\cite{Sankaran2021, Zhang2021, Eyert2008}
Although some ternary alloys, such as MAX phase ceramics, have shown excellent scalability in resistivity and resilience to EM, maintaining a low resistivity in narrow interconnect vias, while reducing process variability remains a significant challenge.
Part of the issue is linked to the need to control both the alloy stoichiometry and the crystal phase within the constrained dimensions.
Therefore, considering ternary or quaternary alloys leads to degrees of complexity and variability that are difficult to master during the integration steps of the interconnects. 

Binary intermetallics can offer more viable solution to reduce some of these variability aspects due to their simpler composition. Their long-range chemical ordering, achieved through proper thermal treatment, allows for strong chemical bonding between neighboring atoms. This bonding increases the cohesive energy, which helps minimize the electromigration process. Many studies already focused on silicides~\cite{Saraswat1985, Sinha1981} and aluminides~\cite{Howell2011, soulie2021, soulie2022, soulie2023, VanTroeye2023, Soulie2024_alcu} but comprehensive reports on thin-film resistivity scaling in the relevant thickness range around or below 10 nm are lacking. As a vast number of combinations to form intermetallics is possible, evaluating their potential as an interconnect metallization layer with an ab initio-based (pre)screening approach is of particular interest.

In this work, we selected a primary metal with low bulk resistivity such as Cu, Al, Ru and Mo (1.7, 2.6, 52 and 70 $\times10^{-8}$ $\Omega$.m,~\cite{haynes2016crc} respectively) to alloy with another transition metals of the periodic table.
It is expected that alloys based on these primary metals would yield low bulk resistivities such as for CuAu.~\cite{zhang2004}
The crystal structures with all different possible combinations and stoichiometries were extracted from the Materials project database.~\cite{Jain2013}
For a robust selection of the most promising binary alloys, we applied a benchmarking methodology similar to that previously reported for elemental metals and ternary MAX phases,~\cite{Dutta2017, Sankaran2021, adelmann2023} namely we compare both $\rho_0 \times \lambda$ product and cohesive energy with reference values of Cu to identify materials that outperform (with lower $\rho_0 \times \lambda$ product and higher cohesive energy) Cu in both aspects.
This approach has successfully identified several alternative candidates to Cu such as Ru, Mo, Co, V$_2$AlC, Cr$_2$AlC.

\section{\label{sec:Methodology}Methodology}

We extracted pre-optimized crystal structures of the most stable Cu, Al, Ru and Mo based binary alloys from the computed phase diagrams provided by the Materials project online database~\cite{Jain2013} and computed their electronic properties using automated first-principles density functional theory (DFT) simulations. Both the structural and electronic properties were determined through the \textsc{\small QUANTUM ESPRESSO} DFT package~\cite{Giannozzi2009} using a combination of plane-waves and pseudopotentials.
The valence electron shells of the elements represented by Garrity–Bennett–Rabe–Vanderbilt (GBRV) pseudopotentials,~\cite{Garrity2014} with a kinetic cutoff energy ranging between 60 and 80 Ry for the truncation of the plane-wave expansion of the wavefunction including a Methfessel-Paxton smearing function with a broadening of 13.6 meV.
The exchange-correlation energy was described within the Perdew–Burke–Ernzerhof generalized gradient approximation.~\cite{Perdew1996} 
The first Brillouin zone was sampled using a discretized Monkhorst–Pack scheme~\cite{Monkhorst1976} based on a regular unshifted ($\Gamma$-point centered) k-point mesh ranging from 40$\times$40$\times$40 to 60$\times$60$\times$60. These ensure a convergence of the total energy within 10$^{-12}$ eV.
In this work, we mainly focused on quantifying the transport properties along different crystallographic directions and for selected different film textures at operating temperatures for interconnect applications, using the methodology introduced in Moors \emph{et al}.~\cite{Moors2022} We ignored other scattering mechanisms such as the electron-phonon interactions and lateral size-effects.~\cite{VanTroeye2023}
The resistivity scalability represented by the $\rho_0\times\lambda$ product is generalized to a tensor form that naturally takes into account both the symmetry group of the material and the anisotropy of the electronic band structures.
This generalization allows for a straightforward comparison between highly anisotropic metals in nanostructures with different lattice orientations and arbitrary transport directions by evaluating the tensor components of $\rho_0\times\lambda$, in this work considering the constant-$\lambda$ approximation.~\cite{Moors2022}
In general, the computed $\rho_0\times\lambda$, assuming $\lambda$ constant, has the following form for the different crystal symmetries in the zero-temperature limit as described in Ref.~\onlinecite{Moors2022}.

\begin{equation}
	\label{eq:rho_lambda}
	\rttensor{\left( \frac{1}{\rho_0\times\lambda} \right)} = \frac{e^2}{\hbar} \sum_n \int \frac{\deriv S_\Fermi^{(n)}}{(2 \pi)^3} \frac{\vecv^{(n)}(\veck) \otimes \vecv^{(n)}(\veck)}{|\vecv^{(n)}(\veck)|^2}
\end{equation}

\noindent with $e$ the electron charge, $\hbar$ the reduced Planck constant, $\vecv^{(n)}(\veck)= \nabla_\mathbf{k} E^{(n)}(\mathbf{k})/\hbar$ the group velocity, $E^{(n)}(\mathbf{k})$ the energy and the integration over the three-dimensional $\veck$-space can be replaced by an integration over the Fermi surface $S_\Fermi^{(n)}$ of each conduction band with index $n$.

It was shown in Ref.~\onlinecite{Moors2022} that the $\rho_0\times\lambda$ tensor has the same symmetry properties as the bulk conductivity tensor. Therefore the $\rho_0\times\lambda$ tensor has the following generalized form:

\begin{equation} \label{eq:rho_lambda_tensor}
	\rttensor{\rho_0\times\lambda} = \begin{pmatrix} 
		(\rho_0\lambda)_{xx} & \rho_0\lambda)_{xy} & (\rho_0\lambda)_{xz} \\
		\rho_0\lambda)_{xy} & (\rho_0\lambda)_{yy} & \rho_0\lambda)_{yz} \\
		(\rho_0\lambda)_{xz} & (\rho_0\lambda)_{yz} & (\rho_0\lambda)_{zz}
	\end{pmatrix}
\end{equation}
The numerical evaluation of $\rho_0\times\lambda$ tensors is computed from the group velocities near Fermi level, which are obtained from the band structure computed using first-principles methods. We previously reported that the anisotropy of the group velocities becomes more pronounced with significant changes in the band structure morphology or the non-uniformity of the Fermi surface.~\cite{Moors2022} 
The $\rho_0\times\lambda$ tensor assumes a constant mean free path. Alternatively, a figure of merit can be constructed under the assumption of constant relaxation time.
In general, the results can be different with this assumption for highly anisotropic materials. However, as shown in Ref.~\onlinecite{Moors2022}, this does not significantly affect the pre-selection of the most promising materials.
Therefore, we proceed with the assumption of constant MFP here.
Additionally, the diagonal $\rho_0\times\lambda$ tensor are sufficient to assess  the sensitivity of metallic resistivity to scaled dimensions for transport along the primary crystallographic directions.
We also evaluate their sensitivity to electromigration (EM) through the evaluation of cohesive energy.
We provide a comprehensive classification of the intermetallics by displaying the diagonal components of $\rho_0\times\lambda$ tensor, while the off-diagonal ones are found to be irrelevant for consideration as Cu replacement,~\cite{Moors2022} and the corresponding cohesive energy, similar to the figure-of-merit as reported in Refs.~\onlinecite{Sankaran2021, adelmann2023, Soulie2024_alcu}.

\section{\label{sec:Results}Results and discussions}

\begin{figure*}[t]
  \centering
  \includegraphics[width=\textwidth]{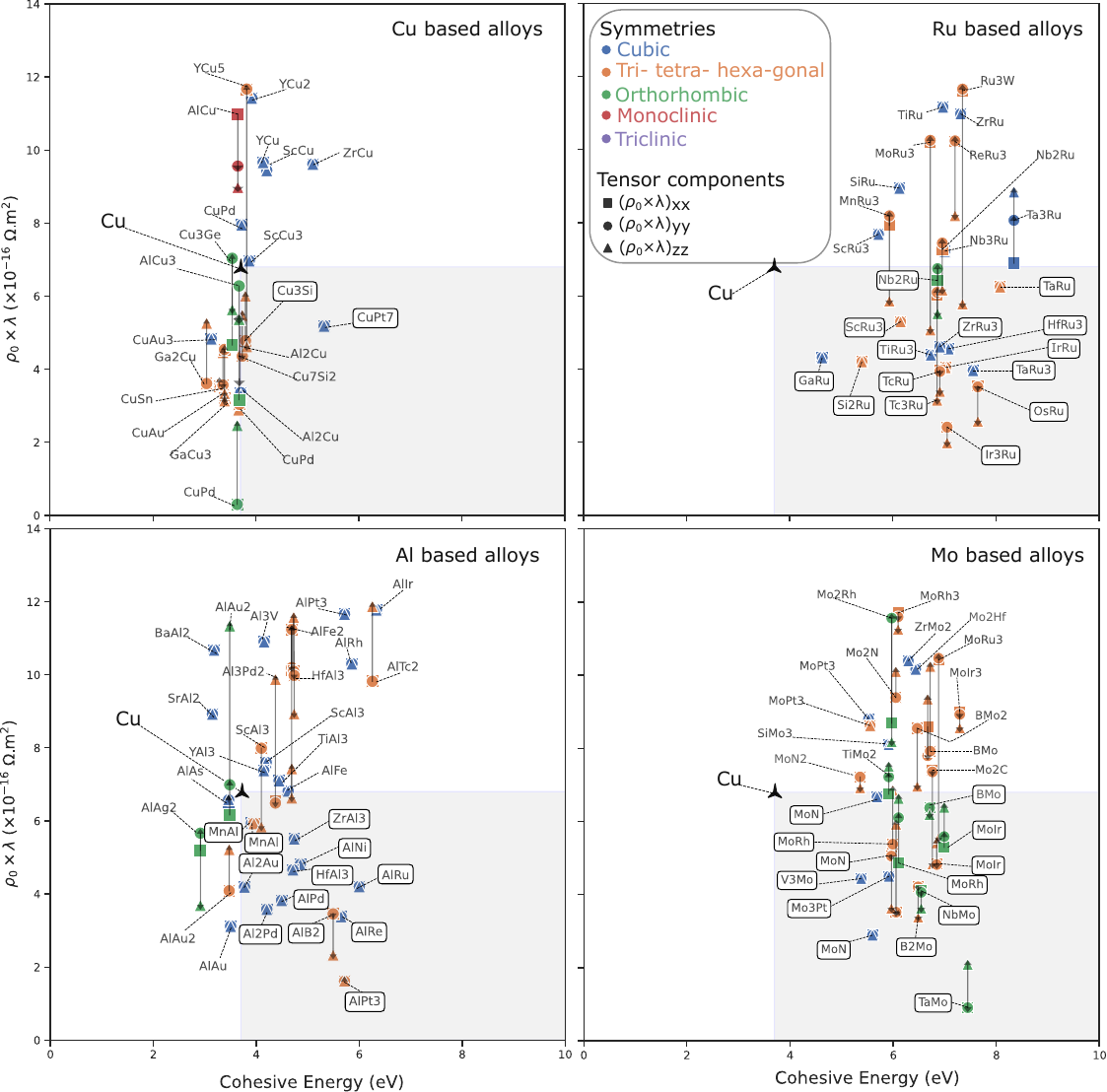}
  \caption{Figures of merit for stable Cu, Al, Ru and Mo binary alloys with Cu as the reference metal (black star). The resistivity scalability potential ($\rho_0\times\lambda$) of binary alloys is represented by both the diagonal tensor components (xx in square, yy in circle, zz in triangle) and the symmetries of binary alloys (cubic in blue, tri-tetra-hexagonal in orange, orthorhombic in green, monoclinic in red and triclinic symmetry in purple) with respect to the resistance to EM (cohesive energy). The light area represent the regions where binary alloys may be expected to have favorable properties with respect to Cu. For clarity, binary alloys outside the gray areas are not encircled.}
  \label{fig:rho0_lambda}
\end{figure*}

In Fig.~\ref{fig:rho0_lambda}, we present the diagonal $\rho_0\times\lambda$ tensor components (xx, yy, zz) for various Cu, Al, Mo and Ru-based binary alloys with negative enthalpy of formation similarly to Ref.~\onlinecite{Soulie2024_alcu}. For the cubic symmetry, only one component of the diagonal tensor is shown,  as all diagonal components of $\rho_0\times\lambda$ are equivalent. For other phases, we report two (three) components for hexagonal, tetragonal, trigonal (and monoclinic, orthorhombic, triclinic), depending on the crystal symmetry, as detailed in Eqs.~\eqref{eq:rho_lambda}-\eqref{eq:rho_lambda_tensor}. Cubic Cu is highlighted as a reference for both $\rho_0\times\lambda$ and cohesive energy (black star at 6.8$\times10^{-16}$ $\Omega$.m$^{2}$ and at 3.6 eV, respectively) in Fig.~\ref{fig:rho0_lambda}, as Cu has been the longstanding standard for nanoscaled interconnect applications.  These dotted lines set the upper and lower limits in our figure-of-merit.
We evaluated more than 270 alloys, but only report 108 of them in Fig.~\ref{fig:rho0_lambda}.  The intermetallics not included typically have $\rho_0 \times \lambda$ values close to twice that of Cu, approximately $12 \times 10^{-16}$ $\Omega \cdot \text{m}^2$.
Out of the 114 candidates, only 41 of them fit within the selection windows defined by Cu. These are listed in Table~\ref{tab:table_1} along with their $\rho_0\times\lambda$ values, cohesive energy and symmetry group together with their corresponding Materials Project (mp-id) identification key. Moreover, the alloys that consistently outperform are those with cubic (c), tetragonal, trigonal hexagonal (tth) and orthorhomic (o) symmetries, due to their lower anisotropic group (Fermi) velocity compared to triclinic or monoclinic symmetries.~\cite{Moors2022} Additionally, many of these intermetallics have excellent scalability and resistance to electromigration. However, despite the excellent scalability potentials of the Cu alloys (see Fig.~\ref{fig:rho0_lambda}), their cohesive energy is relatively similar to that of bulk Cu. This suggests that they will suffer from the same vulnerability to EM as Cu. Moreover, Cu alloys will likely require liners and diffusion barriers to achieve dielectric reliability similar to Cu lines. Interestingly, Al, Mo and Ru intermetallics present excellent alternatives to the Cu alloys.

\begin{table}
    \centering
    \begin{tabular}{ccccc}
    \hline
       \thead{Material} & \thead{$\rho_0\times\lambda_{xx/yy/zz}$} & \thead{Cohesive\\energy} & \thead{MP-id} & \thead{Symmetry\\group}\\
    \hline
        Cu$_{3}$Si & 4.8/4.8/6.0 & 3.8 & mp-972828 & tetragonal\\
        CuPt$_{7}$ & 5.2/5.2/5.2 & 5.3 & mp-12608 & cubic\\
        Al$_{2}$Au & 4.2/4.2/4.2 & 3.8 & mp-2647 & cubic\\
        Al$_{2}$Pd & 3.6/3.6/3.6 & 4.2 & mp-16522 & cubic\\
        AlNi & 4.8/4.8/4.8 & 4.9 & mp-1487 & cubic\\
        AlPd & 3.8/3.8/3.8 & 4.5 & mp-829 & cubic\\
        AlRe & 3.4/3.4/3.4 & 5.6 & mp-10908 & cubic\\
        AlRu & 4.2/4.2/4.2 & 6.0 & mp-542569 & cubic\\
        MnAl & 5.9/5.9/5.9 & 3.9 & mp-12067 & cubic\\
        MnAl & 5.9/5.9/5.9 & 4.0 & mp-771 & tetragonal\\
        HfAl$_{3}$ & 4.7/4.7/4.7 & 4.7 & mp-1007730 & cubic\\
        ZrAl$_{3}$ & 5.5/5.5/5.5 & 4.7 & mp-569775 & cubic\\
        AlPt$_{3}$ & 1.6/1.6/1.6 & 5.7 & mp-1079182 & tetragonal\\
        AlB$_{2}$ & 3.5/3.5/2.3 & 5.5 & mp-944 & hexagonal\\
        Nb$_{2}$Ru & 6.5/6.8/5.7 & 7.0 & mp-1220677 & orthorhombic \\
        ScRu$_{3}$ & 5.5/5.5/5.5 & 6.5 & mp-973022 & tetragonal\\
        TiRu$_{3}$ & 4.7/4.7/4.7 & 7.0 & mp-998947 & cubic\\
        GaRu & 4.6/4.6/4.6 & 5.2 & mp-22320 & cubic\\
        Si$_{2}$Ru & 4.5/4.5/4.5 & 5.9 & mp-7754 & tetragonal\\
        TcRu & 4.3/4.2/3.7 & 7.1 & mp-1217363 & hexagonal\\
        Tc$_{3}$Ru & 6.2/6.2/3.5 & 7.1 & mp-861630 & hexagonal\\
        TaRu & 6.4/6.4/6.4 & 8.1 & mp-1601 & tetragonal\\
        ZrRu$_{3}$ & 4.9/4.9/4.9 & 7.1 & mp-1017544 & cubic\\
        HfRu$_{3}$ & 4.8/4.8/4.8 & 7.3 & mp-1007657 & cubic\\
        IrRu & 4.3/4.3/4.3 & 7.2 & mp-974421 & hexagonal\\
        TaRu$_{3}$ & 4.3/4.3/4.3 & 7.7 & mp-867816 & cubic\\
        OsRu & 3.9/3.9/3.0 & 7.7 & mp-1220023 & hexagonal\\
        Ir$_{3}$Ru & 2.8/2.8/2.5 & 7.2 & mp-974358 & tetragonal\\
        MoN & 6.4/6.4/6.4 & 6.0 & mp-13034 & cubic\\
        MoRh & 5.1/5.1/6.6 & 6.2 & mp-1221419 & hexagonal\\
        MoN & 4.8/4.8/3.4 & 6.2 & mp-13036 & hexagonal\\
        V$_{3}$Mo & 4.2/4.2/4.2 & 5.7 & mp-972071 & cubic\\
        Mo$_{3}$Pt & 4.3/4.3/4.3 & 6.2 & mp-1186016 & cubic\\
        MoN & 2.7/2.7/2.7 & 5.9 & mp-16730 & cubic\\
        BMo & 6.1/6.1/5.9 & 6.8 & mp-999198 & orthorhombic\\
        MoIr & 5.0/5.3/6.1 & 7.1 & mp-11481 & orthorhombic\\ 
        MoIr & 4.6/4.6/5.2 & 6.9 & mp-1221414 & hexagonal\\
        MoRh & 4.6/5.8/6.3 & 6.3 & mp-12595 & orthorhombic\\
        NbMo & 3.9/3.9/3.4 & 6.7 & mp-1220327 & orthorhombic\\
        B$_{2}$Mo & 4.0/4.0/3.2 & 6.6 & mp-960 & hexagonal\\
        TaMo & 0.8/0.8/1.9 & 7.4 & mp-1217895 & orthorhombic\\
    \end{tabular}
    \caption{List of intermetallics that are filtered based on their scalability and resistance to EM (cohesive energy) potentials with respect to Cu as highlighted in Fig.~\ref{fig:rho0_lambda}. The table reports the $\rho_0\times\lambda$ diagonal tensor values (in $\times10^{-16}$ $\Omega$.m$^{2}$), cohesive energy (in eV), materials project key and symmetry group of the identified phases.}
    \label{tab:table_1}
\end{table}

In addition to this initial stringent pruning of intermetallic candidates, it is crucial to pursue the down-selection process using experimental evidence to confirm the existence of the crystal phases identified and their potential in terms of bulk resistivity.
From Table~\ref{tab:table_1}, we could not find any literature reports on the existence of the MoN, TcRu, BMo, NbMo and TaMo. Since the Materials Project database contains crystal structure prototypes, we excluded them from our analysis, as a precautionary measure.
As previously stated, promising intermetallic candidates must also exhibit low bulk resistivity, around 10$^{-8}$ $\Omega$.m. Unfortunately, despite various reports on the crystal textures of these materials, proper characterization of the electrical performances is still lacking. Additionally, computing bulk resistivity using \textit{ab initio} techniques remains very challenging.~\cite{VanTroeye2023} It is hence difficult to draw firm conclusions on the entire list of identified materials. Interestingly, NiAl, RuAl, Sc$_3$Al, AlCu$_3$, Al$_2$Cu, Al$_2$Au have been reported to meet the low bulk resistivity requirement.~\cite{soulie2021, VanTroeye2023, Soulie2024_alcu}
These materials exhibit low bulk resistivity values, which, combined with relatively low $\rho_0\times\lambda$ values, result in MFPs of just a few nanometers. This is significantly smaller than the MFP of Cu (around 40 nm) and comparable to Ru (around 6nm).~\cite{Gall2016,Dutta2017}
Furthermore, the dependence of resistivity on growth orientation and confinement has been explored,~\cite{soulie2021, VanTroeye2023, Soulie2024_alcu} suggesting that NiAl, AlRu and AlCu$_3$ are promising candidates for high conductivity lines in nanoscaled interconnects.
The $\rho_0\times\lambda$ tensor listed in Table~\ref{tab:table_1} from Fig.~\ref{fig:rho0_lambda} are useful for identifying the most promising intermetallics for narrow low resistivity lines.
Therefore, our results can serve as guidelines for experimentally exploring these intermetallics for such applications. However, it is important to note that, depending on the concentration, alloying and metal diffusion can degrade electronic properties at grain boundaries, leading to an increase in resistivity.

\section{\label{sec:Conclusions}Conclusions}
In this work, we investigated potential binary intermetallics as conductor materials for narrow scaled interconnect lines, aiming to replace Cu, whose performance and reliability degrade upon scaling. We computed transport-related $\rho_0\times\lambda$ tensors, which can be used as figure-of-merit for the screening intermetallics scalability without severe degradation of resistivity, along with cohesive energy as a proxy for resistance to EM, using first-principles simulations. 
Among the studied 108 intermetallics, 41 alloys composed of Cu, Al, Ru and Mo primary elemental metals exhibit lower $\rho_0\times\lambda$ tensors and higher cohesive energy than Cu, flagging them as a potential replacement for nano-scale interconnect applications.

\begin{acknowledgments}

This work has been enabled in part by the NanoIC pilot line. The acquisition and operation are jointly funded by the Chips Joint Undertaking, through the European Union’s Digital Europe (101183266) and Horizon Europe programs (101183277), as well as by the participating states Belgium (Flanders), France, Germany, Finland, Ireland and Romania. For more information, visit nanoic-project.eu.
\end{acknowledgments}
	
\label{Bibliography}
\bibliography{binary_alloys}
	
\end{document}